\documentclass[twocolumn,dvipdfm,preprintnumbers,amsmath,amssymb,showpacs]{revtex4}

\usepackage{bm}
\usepackage{graphicx}
\usepackage{dcolumn}
\usepackage{amsmath}%
\usepackage{amssymb}

\begin{document}

\title{Monogamy of Entanglement Measures Based on Fidelity in Multiqubit Systems}

\author{Limin Gao}
\affiliation {College of Physics, Hebei Key Laboratory of Photophysics Research and Application, Hebei Normal University, Shijiazhuang 050024, China}
\author{Fengli Yan}
\email{flyan@hebtu.edu.cn}
\affiliation {College of Physics, Hebei Key Laboratory of Photophysics Research and Application, Hebei Normal University, Shijiazhuang 050024, China}
\author{Ting Gao}
\email{gaoting@hebtu.edu.cn}
\affiliation {School of Mathematical Sciences, Hebei Normal University, Shijiazhuang 050024, China}

\date{\today}

\begin{abstract}
We show exactly that the $\alpha$th power of Bures measure of entanglement and geometric measure of entanglement, as special case of entanglement measures based on fidelity, obey a class of general monogamy inequalities in an arbitrary multiqubit mixed state for $\alpha\geq1$.
\end{abstract}


\maketitle

\emph{Introduction}.---Quantum entanglement as a resource plays a pivotal role in quantum information processing. A characteristic trait of entanglement that differentiates it from classical correlation is known as the monogamy of entanglement (MoE) \cite{1,2}, which means that the more entangled two parties are, the less correlated they can be with other parties. This indicates that the amount of entanglement cannot be shared freely among unrestricted multipartite quantum systems. Since the monogamy of entanglement restricts on the amount of information that an eavesdropper could potentially obtain about the secret key extraction, it plays a crucial role in the security of many information-theoretic protocols, such as quantum key distribution protocols \cite{3,4,5}. MoE also has been widely used in many areas of physics, such as quantum information theory \cite{6}, condensed-matter physics \cite{7} and even black-hole physics \cite{8}.

Monogamy is a very interesting and fundamental property existing only in quantum systems. Determining whether or not a given entanglement measure is monogamous is an important question in the study of monogamy of entanglement. A lot of research work on this topic has been significantly developed over the last two decades.

MoE expresses this principle quantitatively in the form of a mathematical inequality, which was first proved by Coffman, Kundu, and Wootters for three-qubit state in terms of squared concurrence, known as CKW-inequality \cite{1}. Osborne and Verstraete generalized  this three-qubit CKW monogamy inequality  to arbitrary multiqubit systems \cite{9}. Since then, a wide variety of monogamy inequalities for different entanglement measures have been explored extensively \cite{10,11,12,13,14,15,16,17,18,19,20,21,22,23,24,25,26,27,28,29,30,31,32}, such as the entanglement negativity and convex roof extended negativity \cite{10,11,12,13,14,15}, tangle and entanglement of formation \cite{16, 17, 18,19}, Tsallis $q$-entropy and R\'{e}nyi-$\alpha$ entanglement \cite{20,21,22,23,24}. Moreover, Koashi and Winter proved that the one-way distillable entanglement and squashed entanglement satisfy monogamy inequalities for any tripartite systems of arbitrary dimension \cite{25}. Recently, some different kinds of monogamy relations \cite{26,27,28,29,30,31,32} were introduced. The monogamy relations involving other measures of quantum correlations have also been reported \cite{33,34,35,36,37,38,39}.

One class of entanglement measures based on fidelity \cite{40,41,42}, such as Bures measure of entanglement \cite{40,41} and geometric measure of entanglement \cite{42}, which are widely used in multiparticle systems. Recently, we proved analytically that Bures measure of entanglement and geometric measure of entanglement obey a class of general monogamy inequalities in $N$-qubit pure states \cite{46}. However, the case of $N$-qubit mixed states was still open. In this paper, we extend the general monogamy inequalities in terms of Bures measure of entanglement and geometric measure of entanglement for multiqubit pure states to the general monogamy inequalities for mixed states.

\emph{General monogamy inequalities for Bures measure of entanglement and geometric measure of entanglement}.---In order to investigate the monogamy inequalities for the entanglement measures based on fidelity, we need to provide some notations, definitions and lemmas, which are useful throughout this paper.

Consider a quantum system $AB$ consisted of subsystems $A$ and $B$ with respective Hilbert spaces $\mathcal{H}_{A}$ and $\mathcal{H}_{B}$. A pure state $|\psi_{AB}\rangle$ of the quantum system $AB$  in $\mathcal{H}_{A}\otimes \mathcal{H}_{B}$ is called separable if it can be written as $|\psi_{AB}\rangle=|\varphi_{A}\rangle\otimes |\varphi_{B}\rangle$, where $|\varphi_{A}\rangle\in \mathcal{H}_{A}$ and $|\varphi_{B}\rangle\in \mathcal{H}_{B}$.  A mixed state $\sigma_{AB}$ of the quantum system $AB$ is called separable, if it is of the form
\begin{equation}\label{f1}
\begin{aligned}
 \sigma_{AB}= \sum_{i}p_{i} |\varphi^{i}_{A}\rangle\langle\varphi^{i}_{A}|\otimes|\varphi^{i}_{B}\rangle\langle\varphi^{i}_{B}|,
\end{aligned}
\end{equation}
where the $\{p_{i}\}$ forms a probability distribution. We use $S$ to denote the set of separable states, which is a convex set and its extreme points are  pure states.

The fidelity of separability is defined by \cite{43,44}
\begin{equation}\label{f2}
F_s(\rho_{AB})=\max_{\sigma_{AB}\in S} F(\rho_{AB},\sigma_{AB}),
\end{equation}
where the maximum is taken over all separable states $\sigma_{AB}$ in $S$ and $F(\rho_{AB},\sigma_{AB})= (\text{tr}[\sqrt{\sqrt{\rho_{AB}}\sigma_{AB}\sqrt{\rho_{AB}}}])^2$ is Uhlmann's fidelity.

In the following we consider the entanglement measures based on fidelity for two special cases: the first one is Bures measure of entanglement, which can be defined as \cite{40,41}
\begin{equation}\label{f3}
 \begin{aligned}
 E_{\texttt{B}}(\rho_{AB})= \min_{\sigma_{AB}\in S}(2-2\sqrt{F(\rho_{AB},\sigma_{AB})}),
\end{aligned}
\end{equation}
and the second one is geometric measure of entanglement \cite{42}, which is defined as
\begin{equation}\label{f4}
 \begin{aligned}
 E_{\texttt{G}}(\rho_{AB})= \min_{\sigma_{AB}\in S}(1-F(\rho_{AB},\sigma_{AB})).
\end{aligned}
\end{equation}

Then, by employing Eq.(\ref{f2}), Bures measure of entanglement and geometric measure of entanglement can be rewritten in the form
\begin{eqnarray}
E_{\texttt{B}}(\rho_{AB})& = &  2-2\sqrt{F_s(\rho_{AB})}, \label{f5}\\
E_{\texttt{G}}(\rho_{AB})& = &  1-F_s(\rho_{AB}). \label{f6}
\end{eqnarray}

For an arbitrary two-qubit state, the analytical expressions for Bures measure of entanglement and geometric measure of entanglement as a function of the concurrence $C(\rho_{AB})$ are given by \cite{42,43,44}
\begin{eqnarray}
 E_{\texttt{B}}(\rho_{AB})= B(C(\rho_{AB})), \label{f7}\\
 E_{\texttt{G}}(\rho_{AB})= G(C(\rho_{AB})). \label{f8}
\end{eqnarray}
Here
\begin{eqnarray}
B(x)& = & 2-2\sqrt{\frac{1+\sqrt{1-x^{2}}}{2}},\label{f9}\\
G(x)& = & \frac{1-\sqrt{1-x^{2}}}{2}, \label{f10}
\end{eqnarray}
both $B(x)$ and $G(x)$ are monotonically increasing functions in $0\leq x\leq1$.

Next we show the following lemma for $2\otimes d$ systems.

\emph{Lemma 1}. For an arbitrary quantum state $\rho_{AB}$ of $2\otimes d$ systems, we have
\begin{equation}\label{f11}
F_s(\rho_{AB})\leq\frac{1+\sqrt{1-C^2(\rho_{AB})}}{2}.
\end{equation}
Here $C(\rho_{AB})=\text {min}\sum_{j}p_{j}C(|\psi_{j}\rangle_{AB})$, the minimum is taken over all ensemble decompositions of $\rho_{AB}=\sum_{j}p_{j}|\psi_{j}\rangle_{AB} \langle \psi_{j}|$, where $C(|\psi_j\rangle_{AB})$ is the concurrence of quantum state $|\psi_j\rangle_{AB}$, $p_j\geq 0$ and $\sum_j p_j=1$.

\textbf{Proof}. Let us first consider a pure state $|\psi\rangle_{AB}$ of $2\otimes d$ system. Evidently, $|\psi\rangle_{AB}$ has the standard Schmidt decomposition
\begin{equation}\label{f12}
|\psi\rangle_{AB}=\sum\limits_{i=1}^2 \sqrt{\lambda_i}|i_{A}i_{B}\rangle,
\end{equation}
where $\sqrt{\lambda_i}$ are the Schmidt coefficients and $\sum\limits_{i=1}^2 \lambda_i=1$, $|i_{A}\rangle$ and $|i_{B}\rangle$ are the orthonormal basis for the subsystems $A$ and $B$. The concurrence of $|\psi\rangle_{AB}$ is
\begin{equation}\label{f13}
C(|\psi\rangle_{AB})=2\sqrt{\lambda_1 \lambda_2}.
\end{equation}
It has been proved that for bipartite pure state $|\psi\rangle_{AB}$  with the standard Schmidt decomposition Eq.(\ref{f12}) the fidelity of separability is given by \cite{43,45}
\begin{equation}\label{f14}
F_s(|\psi\rangle_{AB})=\max\{\lambda_i\}.
\end{equation}
By calculation, one has
\begin{equation}\label{f15}
F_s(|\psi\rangle_{AB})=\frac{1+\sqrt{1-C^2(|\psi\rangle_{AB})}}{2}.
\end{equation}
Here, we have used the fact that $\sum\limits_{i=1}^2 \lambda_i=1$.

Let us now consider a mixed state $\rho_{AB}$. For an arbitrary quantum state $\rho_{AB}$ and the set $S$ of separable states, the following equation
\begin{equation}\label{f16}
\begin{array}{cl}
F_s(\rho_{AB})&=\max_{\sigma_{AB}\in S} F(\rho_{AB},\sigma_{AB})\\
&=\max_{\{p_{k}, |\psi_{k}\rangle_{AB}\}} \sum_{k}{p_{k}}{F_s}(|\psi_{k}\rangle_{AB})
\end{array}
\end{equation}
holds \cite{43,44}. Suppose that  $\rho_{AB}=\sum_{k}{p_{k}\rho_k}$ with $\rho_k=|\psi_{k}\rangle_{AB}\langle\psi_{k}|$ is the optimal decomposition for $\rho_{AB}$ achieving the maximum of (\ref{f16}). Employing Eq.(\ref{f15}), we can deduce
\begin{equation}\label{f17}
\begin{array}{cl}
F_s(\rho_{AB})& = \sum_{k}{p_{k}}{F_s}(|\psi_{k}\rangle_{AB})\\
\\
& = \sum_{k}{p_{k}}{\frac{1+\sqrt{1-C^2(|\psi_{k}\rangle_{AB})}}{2}}\\
\\
& \leq {\frac{1+\sqrt{1-\sum_{k}p_{k}C^2(|\psi_{k}\rangle_{AB})}}{2}}\\
\\
& \leq \frac{1+\sqrt{1-C^2(\rho_{AB})}}{2}.
\end{array}
\end{equation}
Here we have utilized the concavity of the function $\frac{1+\sqrt{1-x}}{2}$ in the third inequality. Due to that the square of the concurrence  is a convex function on the set of density matrices i.e.,
 $C^2(\sum_{j}p_{j}|\psi_{j}\rangle_{AB}\langle \psi_j|)\leq \sum_{j}p_{j}C^2(|\psi_{j}\rangle_{AB})$ with $p_j\geq 0$ and $\sum_j p_j=1$, and the monotonically decreasing property of the function $\frac{1+\sqrt{1-x}}{2}$, the fourth inequality can be easily obtained. This completes the proof of  inequality (\ref{f11}).   $\hfill\square$

Recalling Lemma in \cite{46}, it yields
\begin{equation}\label{f18}
 \begin{aligned}
B{^{\alpha}}(\sqrt{x^{2}+y^{2}})\geq B{^{\alpha}}(x)+B{^{\alpha}}(y),
\end{aligned}
\end{equation}
and
\begin{equation}\label{f19}
 \begin{aligned}
G{^{\alpha}}(\sqrt{x^{2}+y^{2}})\geq G{^{\alpha}}(x)+G{^{\alpha}}(y),
\end{aligned}
\end{equation}
for $0\leq x, y\leq 1$ such that $0\leq x^{2}+y^{2} \leq 1$ and $\alpha\geq 1$.

In the following, we are ready to present the following theorem, which states that a class of monogamy inequalities for arbitrary multiqubit mixed states can be established using the power of Bures measure of entanglement and geometric measure of entanglement.

\textbf{Theorem}. For an arbitrary $N$-qubit mixed state $\rho_{AB_{1}\cdots B_{N-1}}$ and $\alpha\geq1$, we have
\begin{eqnarray}
  E_{\texttt{B}}^{\alpha}(\rho_{A|B_{1}\cdots B_{N-1}})
 \geq \sum\limits_{i=1}^{N-1}  E_{\texttt{B}}^{\alpha}(\rho_{AB_{i}}),\label{f20}\\
  E_{\texttt{G}}^{\alpha}(\rho_{A|B_{1}\cdots B_{N-1}})
 \geq \sum\limits_{i=1}^{N-1}  E_{\texttt{G}}^{\alpha}(\rho_{AB_{i}}),\label{f21}
\end{eqnarray}
where $\rho_{A|B_{1}\cdots B_{N-1}}$ denotes the state $\rho_{AB_{1}\cdots B_{N-1}}$ viewed as a bipartite state under the partition $A$ and $B_{1}B_{2}\cdots B_{N-1}$, $\rho_{AB_{i}}=\text{tr}_{B_{1}\cdots B_{i-1}B_{i+1}\cdots B_{N-1}}(\rho_{AB_{1}\cdots B_{N-1}})$.

\textbf{Proof}. Note that Eqs. (\ref{f7}) and (\ref{f8}) are not valid for any mixed state $\rho_{AB_{1}\cdots B_{N-1}}$, since the subsystem $B_{1}B_{2}\cdots B_{N-1}$ is not a logic qubit in general. But we can still use the formula (\ref{f5}). Therefore, we have
\begin{equation}\label{f22}
 \begin{aligned}
& E_{\texttt{B}}^{\alpha}(\rho_{A|B_{1}\cdots B_{N-1}}) \\
& \geq  B^{\alpha}(C(\rho_{A|B_{1}\cdots B_{N-1}}))
\\
& \geq B^{\alpha}\left(\sqrt{\sum\limits_{i=1}^{N-1}C^{2}(\rho_{AB_{i}})}\right)\\
& \geq  B^{\alpha}(C(\rho_{AB_{1}}))+B^{\alpha}\left(\sqrt{\sum\limits_{i=2}^{N-1}C^{2}(\rho_{AB_{i}})}\right)\\
& \geq  B^{\alpha}(C(\rho_{AB_{1}}))+\cdots+B^{\alpha}(C(\rho_{AB_{N-1}}))\\
& = E_{\texttt{B}}{^{\alpha}}(\rho_{AB_{1}})+\cdots+E_{\texttt{B}}{^{\alpha}}(\rho_{AB_{N-1}}),
\end{aligned}
\end{equation}
where the first inequality follows immediately from Lemma 1, while the second inequality follows from the monogamy inequality $C(\rho_{A|B_{1}\cdots B_{N-1}})\geq \sqrt{\sum\limits_{i=1}^{N-1}C^{2}(\rho_{AB_{i}})}$ \cite{9} and the monotonically increasing property of the function $B(x)$. The third inequality is due to inequality (\ref{f18}) by letting $x=C(\rho_{AB_{1}})$ and $y=\sqrt{C^{2}(\rho_{AB_{2}})+\cdots+C^{2}(\rho_{AB_{N-1}})}$. The fourth inequality is obtained from the iterative use of inequality (\ref{f18}). Then by Eq. (\ref{f7}), one has the last equality, which completes the proof of inequality (\ref{f20}).

Similarly, inequality (\ref{f21}) follows from inequality (\ref{f11}), the monogamy inequality, the monotonicity of $G(x)$, the iterative use of inequality (\ref{f19}), and Eq. (\ref{f8}).          $\hfill\square$

Both Bures measure of entanglement and geometric measure of entanglement are good measures of entanglement in multiqubit systems for MoE.

\emph{Conclusion}.---In conclusions, we have established a class of monogamy inequalities in an arbitrary multiqubit mixed state based on the $\alpha$th power of Bures measure of entanglement and geometric measure of entanglement for $\alpha\geq1$, which is an important generalization of the former $N$-qubit pure states result \cite{46}. In fact, Theorem can be further refined and become tighter, the results and the proofs can be obtained by following the approach given in Ref. \cite{47}. Our results indicate that Bures measure of entanglement and geometric measure of entanglement are good measures of entanglement for MoE in multiqubit systems. Given the importance of the study on multipartite quantum entanglement, our results not only provide characterizations of multipartite quantum entanglement sharing and distribution among the multipartite systems, but also can provide a rich reference for future work on the study of multipartite quantum entanglement.

\vspace{0.6cm}
\acknowledgments

This work was supported by the National Natural Science Foundation of China under Grant No: 12071110, the Hebei Natural Science Foundation of China under Grant Nos. A2020205014, A2018205125, and  Science and Technology Project of Hebei Education Department  under Grant Nos. ZD2020167, ZD2021066.

\end{document}